\documentclass[final,3p,times,twocolumn]{elsarticle}

\usepackage{amssymb}
\usepackage{amssymb}
\usepackage{url}

\usepackage[figuresright]{rotating}

\usepackage{amsmath,amssymb,amsfonts}
\usepackage{algorithmic}
\usepackage{graphicx}
\usepackage{textcomp}
\usepackage{booktabs}
\usepackage{subfig}
\usepackage{multirow}

\begin{document}

\begin{frontmatter}




\title{Acquire Precise and Comparable Fundus Image Quality Score: FTHNet and FQS Dataset}


\author{Zheng Gong$^{1,*}$, Zhuo Deng$^{1,*}$, Run Gan$^{2}$, Zhiyuan Niu$^{1}$, Lu Chen$^{2}$, Canfeng Huang$^{2}$, Jia Liang$^{2}$, Weihao Gao$^{1}$, \\Fang Li$^{1}$, Shaochong Zhang$^{2,\dagger}$, Lan Ma$^{1,\dagger}$}

\address{$^{1}$ The~Shenzhen~International~Graduate~School, Tsinghua~University\\
$^{2}$ The~Shenzhen~Eye~Hospital\\
$^{*}$ equal contribution \\
$\dagger$ corresponding author
}

\begin{abstract}
\textbf{Objective}: The retinal fundus images are utilized extensively in the diagnosis, and their quality can directly affect the diagnosis results. However, due to the insufficient dataset and algorithm application, current fundus image quality assessment (FIQA) methods are not powerful enough to meet ophthalmologists` demands. 

\textbf{Methods}: In this paper, we address the limitations of datasets and algorithms in FIQA. First, we establish a new FIQA dataset, Fundus Quality Score(FQS), which includes 2246 fundus images with two labels: a continuous Mean Opinion Score varying from 0 to 100 and a three-level quality label. Then, we propose a FIQA Transformer-based Hypernetwork (FTHNet) to solve these tasks with regression results rather than classification results in conventional FIQA works. 

\textbf{Results}: The FTHNet is optimized for the FIQA tasks with extensive experiments.
Results on our FQS dataset show that the FTHNet can give quality scores for fundus images with PLCC of 0.9423 and SRCC of 0.9488, significantly outperforming other methods with fewer parameters and less computation complexity. 

\textbf{Conclusion}: We successfully build a dataset and model addressing the problems of current FIQA methods. Furthermore, the model deployment experiments demonstrate its potential in automatic medical image quality control.

\textbf{Significance}: All experiments are carried out with 10-fold cross-validation to ensure the significance of the results.

\end{abstract}


\end{frontmatter}



\section{Introduction}
\label{sec:introduction}
The retinal fundus image is one of the most commonly used ophthalmology graphics. Many ophthalmologists use fundus images to assist clinical diagnosis of diabetic retinopathy (DR)~\cite{majumder2021multitasking,hua2020convolutional,peng2019deepseenet}, age-related macular degeneration (AMD)~\cite{burlina2016detection}, polypoidal choroidal vasculopathy (PCV)~\cite{cheung2018polypoidal}, and other retinal diseases~\cite{mojab2019deep,liao2019clinical}. The precise diagnosis of eye diseases relies on high-quality(HQ) fundus images. However, fundus images captured with different equipment by ophthalmologists with various levels of experience have large variations in quality. A screening study of 5,575 patients found that about $12\%$ of fundus images are of inadequate quality to be readable by ophthalmologists~\cite{philip2005impact}. Moreover, another study based on UK BioBank also shows that more than 25$\%$ of fundus images need to be of higher quality to allow accurate diagnosis. 
Consequently, low-quality(LQ) fundus images cover a significant percentage of clinical fundus images. According to the experience of ophthalmologists, the common degradation types of LQ fundus images include out-of-focus blur, motion blur, artifact, over-exposure, and over-darkness. The degradation of fundus images may prevent a reliable clinical diagnosis by ophthalmologists or computer-aided systems. Thus, fundus image quality assessment (FIQA) is proposed to help ophthalmologists control the quality of fundus images. The FIQA tasks can be bonded with the collection process of fundus images, which can boost its speed and avoid useless repeats. Moreover, the quality control process in the medical record system can also benefit from FIQA methods. Thus, the research in FIQA is important.

Traditional FIQA methods~\cite{lee1999automatic,lalonde2001automatic,kohler2013automatic} are mainly based on hand-crafted modeling. However, these modeling methods achieve unsatisfactory performance and generality. In recent years, Convolutional Neural Networks(CNNs) have been widely used in image quality assessment(IQA)~\cite{8063957,su2020blindly,sun2022graphiqa,9506075}. Inspired by the success of IQA, CNNs have also been applied to FIQA~\cite{10.1007/978-3-030-32239-7_6,shen2018multi}. Although impressive results have been achieved, CNN-based methods show limitations in capturing long-range dependencies. Recently, the Transformer~\cite{vaswani2017attention} has been introduced into computer vision and outperformed CNN-based methods in many tasks. The Multi-head Self-Attention(MSA) in the Transformer exhibits excellent performance modeling non-local similarity and long-term dependencies. This advantage of the Transformer may address the limitations of CNN-based models.

Meanwhile, most present FIQA methods are data-driven, which means the performance of these methods relies on the quality of the data. Unfortunately, a professional and available clinical benchmark has yet to be explored. \textbf{Firstly}, some datasets~\cite{raj2019fundus,strauss2007image,10.1007/978-3-030-32239-7_6} treat the FIQA task as a classification task rather than a regression task. The fundus images are roughly divided into several categories, such as ``Good", ``Usable", and ``Bad" quality. The roughly labeled images can lead to images of prominently different quality classified in the same category. \textbf{Secondly}, when a quality scale is used, some datasets do not consider that the image quality of different fundus regions has different effects on clinical diagnosis. For example, the influence of optic disc regions on the fundus image quality should be more significant than that of edge regions. \textbf{Thirdly}, many works train and test on their private datasets, which are labeled using their own standard and are not publicly available. Therefore, it is not convenient to benchmark the performance of the FIQA methods.

\begin{figure*}[htbp]
\centering
\includegraphics[width=0.95\textwidth]{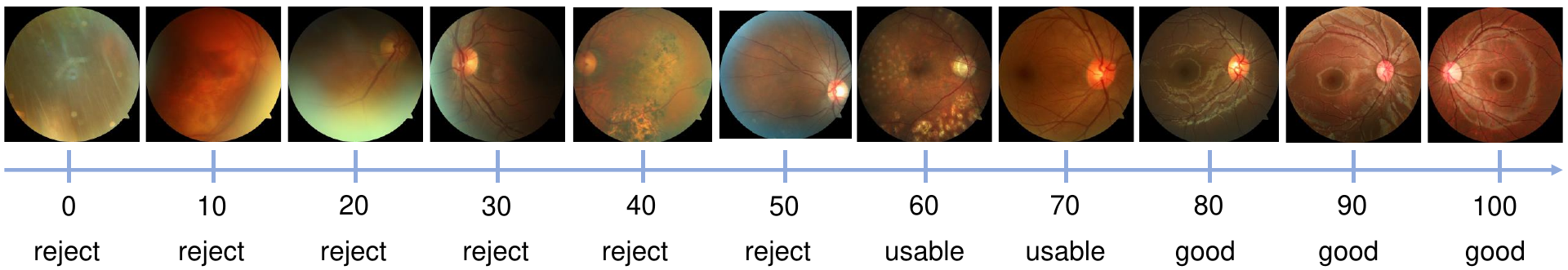}
\caption{Exapmles of the Fundus Quality Score (FQS) dataset. In the FQS dataset, each fundus image has two labels: a three-level classification label (good, reject, and usable) and a continuous MOS varying from 0 to 100. Our FQS dataset covers the most common degradation types in clinical diagnoses, such as out-of-focus blur, haze, uneven illumination, and over-darkness.}
\label{figexample}
\end{figure*}

In this paper, we set out to address the limitations of algorithms and datasets in FIQA. To begin with, we establish a new clinically acquired dataset, Fundus Quality Scores (FQS), including 2246 fundus images with continuous mean opinion scores (MOSs) ranging from 0 to 100 and three-level labels (``Good", ``Reject", and ``Usable"). Based on this dataset, we propose a novel method for FIQA, namely the FIQA Transformer-based HyperNetwork (FTHNet). The proposed FTHNet consists of four parts: the Transformer Backbone, the Distortion Perception Network, the Parameter Hypernetwork, and the Target Network. Specifically, the Transformer Backbone is built up by Basic Transformer Blocks (BTBs). The self-attention mechanism in BTBs can capture the non-local self-similarity and long-term dependencies, which are the main limitations of existing CNN-based methods. The proposed Distortion Perception Network can collect distortion information in different resolutions. We introduce the Parameter Hypernetworks, which can dynamically generate weights and biases according to fundus image contents. Furthermore, the Target Network receives the weights and biases and predicts fundus image quality scores. This proposed method is supposed to give prediction scores consistent with ophthalmologists' experience and perception. 

Our contributions can be summarized as follows:
\begin{itemize}
    \item We establish a new clinical dataset, FQS, to evaluate FIQA algorithms. This is the first professional and available FIQA dataset with both continuous MOSs and three-level classification labels.
    \item We propose a novel FIQA model with the Transformer-based hypernetwork, FTHNet. It is the first attempt to introduce the Transformer aligning with hypernetwork to FIQA tasks.
    \item  Experimental results demonstrate that our FTHNet significantly outperforms current algorithms in the FIQA tasks with fewer parameters and less computation complexity, and thus has excellent potential in real-time diagnosis assistance.
\end{itemize}

\section{Material and methods}

\subsection{Fundus Quality Score}

This subsection introduces our clinical dataset, Fundus Quality Score (FQS). It contains 2246 fundus images with a spatial size of $2304 \times 1728$. In FQS, each image has two labels, a three-level classification label (good, reject, and usable) and a continuous Mean Opinion Score (MOS) varying from 0 to 100 with a minimum gap of 1.

\subsubsection{Data Collecting}
We select 2246 fundus images with typical degradation or pathological features from over 10,000 eye instances. Specifically, 92\% of selected images are from the usual clinical diagnosis across all ages and genders, and the other 8\% are from teens' myopia screening, which has higher image quality and fewer pathological features. These images can cover most retinal photographing formations of eye hospitals. 

The collection process of our FQS is approved and supervised by Shenzhen Eye Hospital. The fundus images are captured by ophthalmologists using a ZEISS VISUCAM200 fundus camera or a Canon fundus camera, which are the mainstream products of fundus cameras. Sensitive information, such as names and diagnosis results, is deleted from the beginning of data collection. We also deleted the EXIF metadata of fundus images.


\begin{figure*}[htbp]
\centering
\includegraphics[width=0.85\textwidth]{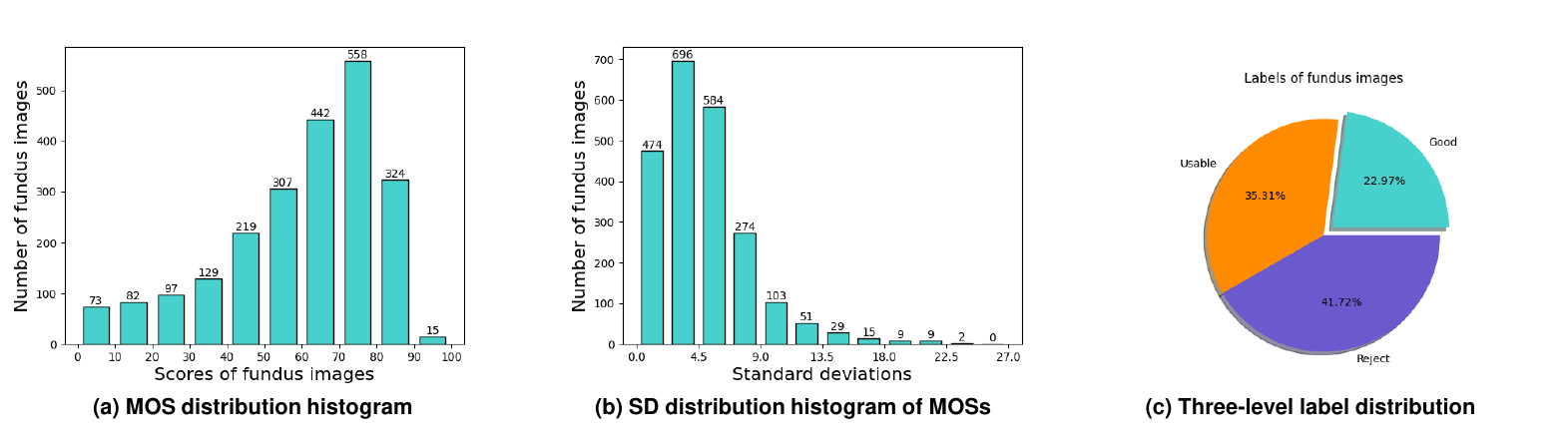}
\caption{The statistic information of our FQS. (a) The MOS distribution histogram. Most of the MOSs are distributed between 60 and 80, which is consistent with actual clinical experience, and images of either extremely high or low quality are rare. (b) The standard deviation distribution histogram of MOSs. Half of the images have standard deviations under 4.34. Low SDs indicate that, though the opinion scores are given independently, the scoring criteria are consistent. (c) The three-level label distribution. The numbers of `Good', `Usable', and `Reject' are 516, 793, and 937. There are more `Reject' images to cover more degradation types.}
\label{figscore} 
\end{figure*}

\subsubsection{Labelling}
In our FQS, we propose a new and complete fundus image quality scale including a three-level classification label (``Good", ``Reject", and ``Usable") and a continuous MOS varying from 0 to 100 with a minimum gap of 1. An example of the quality scale can be seen in Fig.~\ref{figexample}.

Unlike natural IQA, the fundus image quality evaluation needs more professional consideration of the influence of image quality on clinical diagnosis. Therefore, we invite 3 ophthalmologists and design a set of rigorous and scientific labeling processes.

\textbf{Firstly}, we set up the reference set, including 330 fundus images with three-level classification labels (``Good", ``Reject", and ``Usable") and MOS. The reference set is constructed by an experienced ophthalmologist based on clinical experience. Then, the images in each category are further classified into five levels by their degree of degradation (extent of the out-of-focus blur, haze, uneven illumination, etc.). The scores of these images are further altered by assessing the following questions to determine their final MOSs:
\begin{itemize}
    \item [(1)] Whether the degree of fundus image degradation is serious compared to the images in the same category.
    \item [(2)] Whether the structural information that fundus images provide (blood vessel shape and quantity) is comprehensive.
    \item [(3)] Whether the pathological feature is clear enough for clinical diagnosis.
    \item [(4)] Whether the critical position (macula, optic disk, etc.) is visible and clear.
\end{itemize}

\textbf{Secondly}, three experienced ophthalmologists gave their opinion scores for 2246 images under the guidance of the reference set. In addition, we also invited three other junior ophthalmologists. To make the labeling process more convenient, we built the software FundusMarking (shown in Appendix, Fig.\ref{figexe}). This software has a straightforward GUI and can be easily modified to build any new IQA datasets. Note that scores will be discussed and adjusted if there is a significant difference between the opinion scores of the same image. \textbf{Finally}, we obtain the MOS by the weighted average of six independent scores:
\begin{small}
\begin{equation}
    \mathbf{MOS} = \lambda_{1}\sum_{i=1}^{3}\mathbf{O}_{i} + \lambda_{2}\sum_{i=1}^{3}\mathbf{Oj}_{i},
\label{eq9}
\end{equation}
\end{small}
where $\mathbf{O}_i$ and $\mathbf{Oj}_i$ denote the opinion scores of experienced ophthalmologists and junior ophthalmologists. $\lambda_1$ and $\lambda_2$ represent the weights of junior ophthalmologists and experienced ophthalmologists, which are set to 0.11 and 0.22.

\subsubsection{Statistic Information}

\textbf{MOS Distribution Histogram} Fig.\ref{figscore} (a) depicts the distribution of MOS. It can be observed that most of the MOSs are distributed between 60 and 80. The number of images with MOS distributed between 70 and 80 is the largest, and the number of images from 90 to 100 is the lowest, with only 15. The MOS distribution in our FQS is natural and consistent with actual clinical experience. Affected by the equipment and patient coordination, it is difficult to obtain very high-quality fundus images (MOS: 80 or above) in actual clinical collecting. Meanwhile, retinal fundus images of extremely low quality are also rare. The quality of most clinical fundus images is in the upper middle of the score scale.

\textbf{Three-level Label Distribution} The distribution of the three-level classification label is shown in Fig.\ref{figscore} (c). 
Since the HQ images are similar, but the LQ images are different in numerous aspects, there are more ``Reject" images in the FQS dataset to cover more degradation types.
The numbers of fundus images labeled ``Good", ``Usable", and ``Reject" are 516, 793, and 937, respectively. 
The distribution of the three-level classification label is consistent with the MOSs, which is common in actual clinical fundus image collecting.



\textbf{Standard Deviation Distribution Histogram} Since the MOS of each fundus image consists of multiple independent opinion scores from several ophthalmologists, it is necessary to evaluate the uniformity of these six scores. We calculate the standard deviation (SD) of each image and make the following SD distribution histogram (Fig. \ref{figscore} (b)). About a quarter of the SDs are less than 2.52, half of the SDs are under 4.34, and three-quarters are under 6.37. Thus, the distribution of SD indicates the consistency of opinion scores and proves that prior knowledge of the labeling process is consistent between opinion givers.

\begin{figure*}[htbp]
\centering
\includegraphics[width=0.95\textwidth]{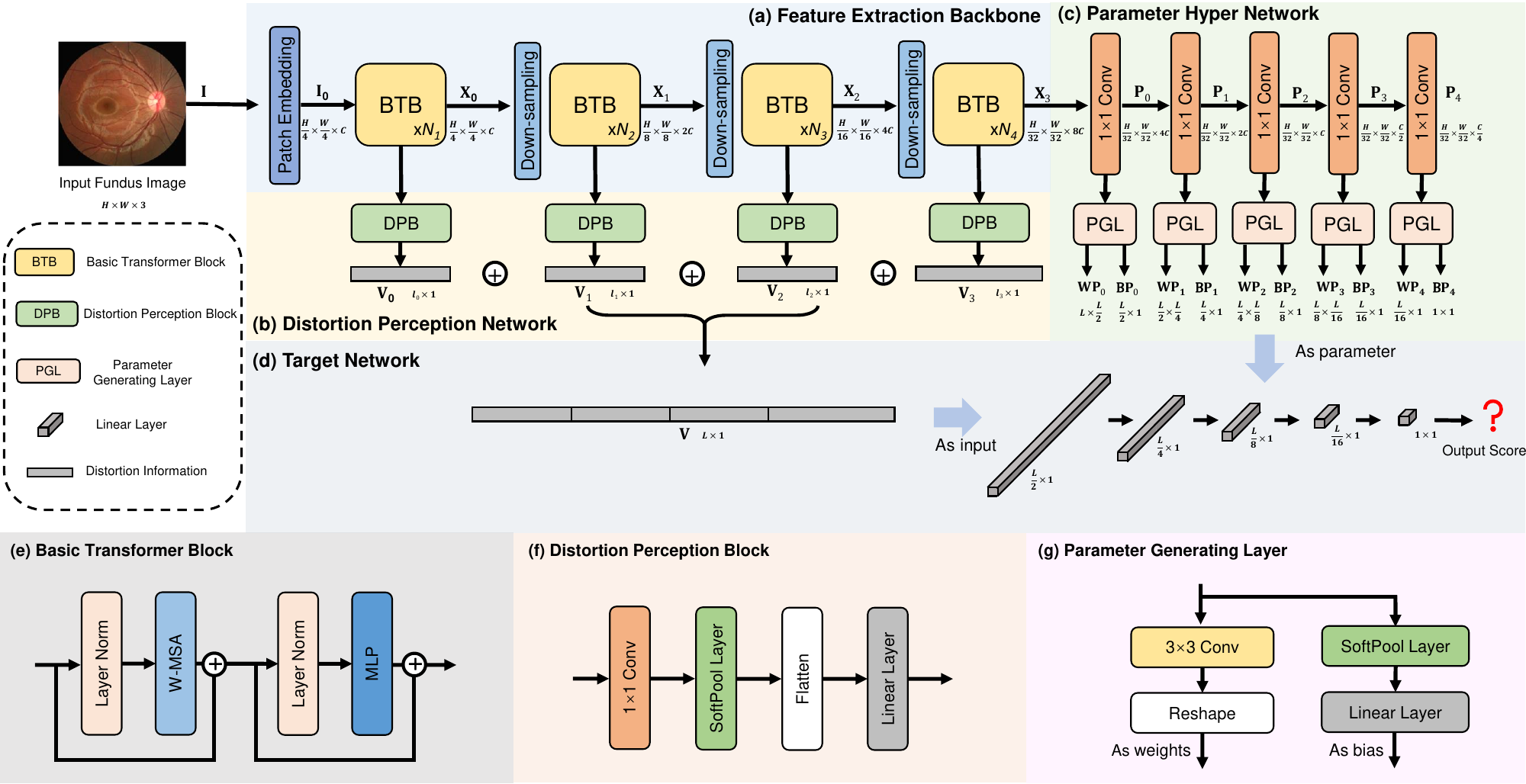}\caption{Architecture of our FTHNet. (a) The Transformer Backbone includes a patch embedding layer and four feature extraction stages. (b) The Distortion Perception Network is designed to extract distorted information. (c) The Parameter Hypernetwork comprises five parameter-generating layers. (d) The Target Network contains five linear layers to predict the fundus image quality scores. (e) The structure of the Basic Transformer Block. (f) The distortion perception block extracts the distortion information from the feature maps in different resolutions. (g) Each parameter-generating layer includes two branches to generate weight and bias parameters.}

\label{fig1}
\end{figure*}

\subsection{method}
\subsubsection{Overall Architecture}
The architecture of FTHNet is shown in Fig.\ref{fig1}, which consists of 4 parts: the Transformer Backbone, the Distortion Perception Network, the Parameter Hypernetwork, and the Target Network.

The Transformer Backbone contains a patch embedding layer and four stages in different resolutions. Each stage comprises several Basic Transformer Blocks (BTBs) and a downsampling layer. \textbf{Firstly}, assuming an input fundus image, $\mathbf{I}_{in} \in\mathbb{R}^{H\times W \times 3} $, the Transformer Backbone exploits a non-overlapping patch embedding layer consisting of a $ 4\times4 $ convolution (\emph{conv}) to extract shallow feature $\mathbf{I}_{0} \in\mathbb{R}^{\frac{H}{4}\times \frac{W}{4} \times C} $. \textbf{Secondly}, 4 stages are used for feature extraction on $\mathbf{I}_{0}$. We adopt a $4 \times 4$ \emph{conv} with stride 2 as the downsampling layer to the feature maps and double the channel dimension. Thus, the feature of the \emph{i}-th stage is denoted as $\mathbf{X}_{i} \in \mathbb{R}^{\frac{H}{4\times 2^i} \times \frac{W}{4\times2^i} \times 2^{i}C}$. Here, \emph{i} = 0, 1, 2, 3 indicates the four stages. 

We design the Distortion Perception Network to extract distorted information, as shown in Fig. \ref{fig1}(b). Each stage input is processed in the Distortion Perception Block (DPB). The input feature maps are $\mathbf{X}_{i}, i=0,1,2,3$. Then, we get four distorted information vectors through the DPB. Finally, distorted information vectors are concatenated to obtain the semantic vector $\mathbf{V} \in \mathbb{R}^{L \times 1}$.

The parameters of the Target Network are generated from the Parameter Hypernetwork shown in Fig.\ref{fig1}(c). We design five stages for processing the parameters: one $1\times1$ \emph{conv} layer for channel merging and one Parameter Generating Layer (PGL). The Parameter Hypernetwork gets input directly from the final stage of the Transformer Backbone. The input feature map is denoted as $\mathbf{X}_{3} \in\mathbb{R}^{\frac{H}{32} \times \frac{W}{32} \times 8C}$. We adopt a $1\times1$ \emph{conv} layer to merge channels by half. After channel merging, the feature of  the \emph{i}-th stage is denoted as $\mathbf{P}_{i} \in\mathbb{R}^{\frac{H}{32} \times \frac{W}{32} \times \frac{8c}{2^i}}$. Here, \emph{i} = 0, 1, 2, 3, 4 indicates the five stages. Then, the Parameter Hypernetwork exploits the PGL to generate the weights and biases of the linear layers for the Target Network. The weights and bias are denoted as $\mathbf{WP}_{i}$and $\mathbf{BP}_{i}$.

The Target Network consists of five linear layers. The input is the $\mathbf{V}$ from the Distortion perception Network. The weights and biases of the Target Network are from the Parameter Hypernetwork. We utilize five linear layers to generate the predicted fundus image quality scores. 


In the implementation, we change the patch embedding channel $C$ and the combination ($N_1, N_2, N_3, N_4$) in the Transformer Backbone to establish two different FTHNet models: FTHNet-S: (2,4,6,2), $C$:32; FTHNet-L: (2,4,6,2), $C$:64.

\subsubsection{Basic Transformer Block}

The emergence of Transformer provides an alternative to address the limitations of CNN-based methods in modeling non-local self-similarity and long-range dependencies. However, the computational cost of the standard global Transformer is quadratic to the spatial size of the input feature (HW). Therefore, to avoid this problem, we apply the transformer blocks with the Window-based Multi-head Self-Attention (W-MSA)~\cite{liu2021swin} in the Transformer Backbone. The computational complexity of W-MSA is linear to the spatial size, which is more wieldable than standard global MSA. We also add the alternate window-shifting operation in the BTB to introduce cross-window connections. The BTB consists of one W-MSA, one Multilayer Perceptron (MLP), and two normalization layers, as shown in Fig.~\ref{fig1}(e). BTB can be formulated as eq.\ref{eq:wmsa}.
\begin{small}
\begin{equation}
    \begin{aligned}
   & \mathbf{F}^{\prime} = \text{W-MSA}(\text{LN}(\mathbf{F}_{in}))+\mathbf{F}_{in},  \\
   & \mathbf{F}_{out} = \text{MLP}(\text{LN}(\mathbf{F}^{\prime}))+\mathbf{F}^{\prime},
    \end{aligned}
    \label{eq:wmsa}
\end{equation}
\end{small}
where $\mathbf{F}_{in}$ represents the input feature maps of a BTB. $\text{LN}(\cdot)$ represents the layer normalization. $\mathbf{F}^{\prime}$ and $\mathbf{F}_{out}$ denote the output feature of W-MSA and MLP respectively.

where Concat$(\cdot)$ denotes the concatenating operation and  $\mathbf{W}^{O}\in\mathbb{R}^{{C \times  C}} $ are learnable parameters. We reshape $\mathbf{X}^{i}_{o}$ to obtain the output window feature map $\mathbf{X}^{i}_{out} \in\mathbb{R}^{{L \times L \times C }}.$ Finally, we merge all the patch representations $\{\mathbf{X}_{out}^{1}, \mathbf{X}_{out}^{2},\mathbf{X}_{out}^{3}, \cdots , \mathbf{X}_{out}^{N} \}$ to obtain the output feature maps $\mathbf{X}_{out} \in\mathbb{R}^{{H \times W \times C }}$.

\textbf{Multilayer Perception}

The Multilayer Perception(MLP) in the BTB resembles the most utilized methods in Transformers, consisting of a linear layer with a GELU activation, two dropout layers, and another linear layer.

\subsubsection{Distortion Perception Block}

We design the DPB in the Distortion Perception Network to extract the distortion information from the feature maps in different resolutions. 
Distortions of the fundus images are widely distributed on different scales. For example, spots and flares affect only some small areas; over-darkness and over-exposure influence the whole image. Rather than dealing with the final feature map, the DPBs get inputs from four backbone stages in different resolutions. 

As depicted in Fig.~\ref{fig1}(f), the DPB consists of one $1\times 1$ \emph{conv} layer, one SoftPool\cite{stergiou2021refining} layer, and one linear layer. We adopt the $1\times1$ \emph{conv} and SoftPool to merge the channels and to downscale the spatial size of feature maps, respectively. The DPB can be formulated as follows:
\begin{small}
\begin{equation}
    \begin{aligned}
    & \mathbf{X}^{\prime} = \text{SoftPool}(1\times1 ~ \text{Conv}(\mathbf{X})), \\
    & \mathbf{V} = \text{Linear}(\text{Flatten}(\mathbf{X}^{\prime})),
    \end{aligned}
\end{equation}
\end{small}
where $\mathbf{X} \in \mathbb{R}^{H \times W \times C}$ are the input feature map of a DPB. $\text{SoftPool}$, $1\times1 ~ \text{Conv}$, $\text{Linear}$, and $\text{Flatten}$ represents the SoftPool layer, the $1\times 1$ \emph{conv} layer, the linear layer, and flatten operation, respectively. $\mathbf{X}^{\prime} \in \mathbb{R}^{\frac{H}{12} \times \frac{W}{12} \times \frac{C}{8}}$ denotes the output feature. $\mathbf{V} \in \mathbb{R}^{l \times 1}$ is the distortion information vector.

\subsubsection{Parameter Generating Layer}

The parameters of the Target Network are processed by the Parameter Generating Layer (PGL) shown in Fig.~\ref{fig1}(g). The PGL consists of two branches, one for generating the weight parameters and another for generating the bias parameters. The weight branch is formulated as follows:
\begin{small}
\begin{equation}
    \mathbf{WP} = \text{Reshape}(3\times3 ~ \text{Conv}(\mathbf{P}))
\end{equation}
\end{small}
Where $\mathbf{P}$ denotes the input feature map of PGL. $\mathbf{WP}$ represents the weight parameter. $3\times3 ~ \text{Conv}$ and $\text{Reshape}$ represent the $3 \times 3$ \emph{conv} layer and reshape operation respectively. 

Similarly, the bias branch can be formulated as follows:
\begin{small}
\begin{equation}
    \mathbf{BP} = \text{Linear}(\text{SoftPool}(\mathbf{P}))
\end{equation}
\end{small}
where $\mathbf{BP}$ represents the weight parameter. $\text{Linear}$ and $\text{SoftPool}$ denote the linear layer and SoftPool layer respectively.

Though this process seems complicated, it can directly generate the parameter matrix with desired shape and length if we calculate the output channel using $\frac{Channel_{in}\times H\times W}{\frac{L}{2^{i-1}} \times \frac{L}{2^i}}$, where $H\times W$ represents the size of the feature map, and $L$ represents the input length of Target Network in the $ith$ stage.

\subsubsection{Loss Function}
We choose the smooth L1 loss\cite{7410526} as our model's loss function. To be specific, the smooth L1 loss is formulated as 

\begin{small}
\begin{equation}
    \mathcal{L}_{smoothL1} = \begin{cases}
    |\mathbf{y}-\mathbf{y}^{\prime}|-0.5, & otherwise \\
0.5\times(\mathbf{y}-\mathbf{y}^{\prime})^2, & -1<\mathbf{y}-\mathbf{y}^{\prime}<1 \\
\end{cases}
\end{equation}
\end{small}
where $\mathbf{y}$ denotes the ground-truth fundus image quality score, $\mathbf{y}^{\prime}$ denotes the predicted score.

\section{Results}

\subsection{Implementation Details}
The train, test, and validation subsets are split proportionately 80\%, 15\%, and 5\% randomly in each round, and 10-round cross-validation is applied in training. During the training procedure, Fundus images are resized to $384 \times 384$. The Adam~\cite{kingma2014adam} optimizer is adopted. We also apply data augmentation consisting of horizontal/vertical flip and random crop. The learning rate is set to $0.5 \times 10^{-4}$ with linear annealing, and the batch size is set to 16. We use PyTorch 1.9 and CUDA 11.2. Every model is trained for 120000 iterations with a warming-up of 1000 iterations, equivalently 853 epochs. It takes about 12 hours to use an NVIDIA RTX3090 GPU to finish the training process. 

We chose root mean square error (RMSE), Pearson correlation coefficient (PLCC), and Spearman's rank correlation coefficient (SRCC) as the metrics to evaluate the performance of our model.

The RMSE in eq.~\ref{eqrmse} shows the deviation between the prediction and target values. 
The lower the RMSE, the more robust the model in prediction.
\begin{equation}
RMSE = \sqrt{\frac{1}{n}\sum_{i=1}^{n}(y^{\prime}_{i}-y_{i})^2}
\label{eqrmse}
\end{equation}
where $y_{i}$ denotes the ground-truth MOS, $y^{\prime}_{i}$ denotes the predicted MOS.

The PLCC and SRCC are both standard metrics applied in the IQA works. The PLCC shown in eq.\ref{eqplcc} varies from -1 to 1 and shows the correlation between prediction and target values. The higher PLCC means the model's predictions are closer to the images' actual scores.
\begin{equation}
PLCC = \frac{\sum_{i}(y^{\prime}_i-\Bar{y}^{\prime})(y_i-\Bar{y})}{\sqrt{\sum_i(y^{\prime}_i-\Bar{y}^{\prime})^2\sum_i(y_i-\Bar{y})^2}}
\label{eqplcc}
\end{equation}
where $\Bar{y}^{\prime}$ represents the average value of $y^{\prime}_{i}$ and $\Bar{y}$ represents the average value of $y_{i}$. 

The SRCC in eq.\ref{eqsrcc} varies from -1 to 1 and measures the monotonicity of the models' prediction. 
\begin{equation}
SRCC = 1 - \frac{6\sum^{N}_{i=1}{d_{i}^2}}{N(N^2-1)}
\label{eqsrcc}
\end{equation}
where $d_{i} = y^{\prime}_{i} - y_{i}$ denotes the difference between $y^{\prime}_{i}$ and $y_{i}$.

\begin{table}
	\footnotesize
	\centering	
	\caption{Quantitative comparisons with SOTA algorithms on our FQS.}
	\scalebox{1}{
			\begin{tabular}{c c c c c }
				\toprule
				 Method & SRCC & PLCC & RMSE  & Params(M)  \\
				\midrule
				TRIQ~\cite{9506075} & 0.4153& 0.4327& -- &23.68  \\
				DeepIQA~\cite{8063957} & 0.2964 & 0.2798 & -- &6.29  \\
				HyperIQA~\cite{su2020blindly} & 0.9351 &0.9305 & -- & 28.28 \\
				GraphIQA~\cite{sun2022graphiqa}  &0.9280  & 0.9355  & -- &51.52\\
				BRISQUE~\cite{6272356} &0.9020  & 0.9220 & 8.039   & -- \\
				ILINIQE~\cite{7094273} &0.6550 & 0.7191&35.46  &--  \\
				BIQI~\cite{5432998} & 0.5788 & 0.5492 & 76.2& --\\
				DIVINE~\cite{5756237} & 0.1566 & 0.1686 & 43.19 & --\\
				\midrule
				FTHNet-S~(Ours) & 0.9330 & 0.9435 & 7.118 & \textbf{5.662}  \\
                \textbf{FTHNet-L~(Ours)} & \textbf{0.9358} & \textbf{0.9442} & \textbf{7.024}& 14.88\\
				\bottomrule
	\end{tabular}}
	\label{tab:sota}
\end{table}

\subsection{Comparisons with State-of-the-Art Methods}

We compare our FTHNet with several SOTA methods, including four model-based methods (BRISQUE\cite{6272356}, ILINIQE\cite{7094273}, BIQI\cite{5432998}, and DIVINE\cite{5756237}) and four deep learning methods (DeepIQA\cite{8063957}, HyperIQA\cite{su2020blindly}, GraphIQA\cite{sun2022graphiqa}, and TRIQ\cite{9506075}). 


As shown in the Tabel.\ref{tab:sota}, The model-based methods mentioned above can merely reach the performance of FTHNet, though these methods are retrained on our FQS dataset. Since the model-based methods do not have network parameters and computing flops, the corresponding data is not shown in the table.

The IQA methods had not been applied to FIQA tasks before this work, and we retrained HyperIQA, GraphIQA, and TRIQ for their best performance in the comparison study. The quantitative comparisons on our FQS dataset are shown in Table.\ref{tab:sota}. Our best model, FTHNet-L, achieves 0.0133, 0.0183, 0.5161, and 0.6690 improvements in PLCC compared to GraphIQA, HyperIQA, TRIQ, and DeepIQA. Meanwhile, FTHNet-L has 0.0143, 0.0072, 0.5270, and 0.6459 better performance in SRCC, respectively. These improvements of FTHNet-L are significant enough for IQA tasks.

Furthermore, although FTHNet-L is already smaller than most other models, our lightweight model, FTHNet-S, outperforms other methods by 0.008, 0.013, 0.5108, and 0.6637 in PLCC, respectively, while requiring 5.67 M Params.


\begin{table*}
	\footnotesize
	\centering	
	\caption{Ablation study of Transformer Backbone, window shifted operation(WSO), loss function, downsampling structure, and hypernetwork structure.}
	\scalebox{1.2}{
			\begin{tabular}{c c c c c c c}
				\toprule
				 Type & Method & SRCC & PLCC & RMSE  & Params(M)&FLOPS \\
				\midrule
				\multirow{4}{*}{Backbones} & ResNet\cite{he2016deep} & 0.1672 & 0.1630 & 39.24 & 27.97& 12.70 G\\
                & ConvNeXt-L\cite{liu2022convnet} & 0.8095 & 0.8081 & 12.19 & 55.25& 26.30 G\\
                & MSG Transformer\cite{fang2021msg} & 0.9241 & 0.9277 & 7.615 & 33.29& 16.02 G\\
                &\textbf{BTB} & \textbf{0.9358} & \textbf{0.9442} & \textbf{7.024}& \textbf{14.88}& \textbf{6.044 G}\\
                \midrule
                \multirow{2}{*}{WSO}& w/o & 0.9263 & 0.9405 & 7.356 & 14.88 & 6.044 G \\
                 & with & \textbf{0.9358} & \textbf{0.9442} & \textbf{7.024}& 14.88 & 6.044 G\\
                \midrule
                \multirow{4}{*}{Loss}& $\mathcal{L}_1 $& 0.93245 & 0.9439 & 6.716 & 14.88 & 6.044 G \\
                &$\mathcal{L}_2$ & \textbf{0.9363} & 0.9389 & 7.028& 14.88 & 6.044 G\\
               &$\mathcal{L}_1 + \mathcal{L}_2$ & 0.93405 & 0.9408 & 6.914 & 14.88 & 6.044 G\\
                &$\mathcal{L}_{smoothL1}$ & 0.9345 & \textbf{0.9447} & \textbf{6.581} & 14.88  & 6.044 G\\
                \midrule
                \multirow{2}{*}{Downsampling Structure}& Direct & \textbf{0.9357} & 0.9426 & \textbf{6.978} & 0.572 & 82.39 M\\
                &Stepwise & 0.9354 & \textbf{0.9437} & 6.988& \textbf{0.393} &\textbf{56.67 M}\\
                \midrule
                \multirow{2}{*}{hypernetwork}& w/o  & 0.6092 & 0.6077 & 20.93 & \textbf{12.28} & \textbf{5.06 G}\\
                & with & \textbf{0.9358} & \textbf{0.9442} & \textbf{7.024} & 14.88 & 6.044 G\\
				\bottomrule
	\end{tabular}}
	\label{tbbb1}
\end{table*}

\subsection{Ablation Study}
\subsubsection{Transformer Backbone}
To analyze the effect of the Transformer Backbone, we compare our BTB with three solid Transformer Backbones in Table.~\ref{tbbb1}, including two CNN-based backbones (ResNet\cite{7780459} and ConvNeXt\cite{liu2022convnet}) and one Transformer-based backbone(MSG-Transformer\cite{fang2021msg}) with other structure of FTHNet unchanged. It can be observed in the table that our proposed BTB achieves the best performance.

\subsubsection{Window Shift Operations}
The window shift operations are designed to augment the information exchange between adjacent windows in the BTBs. We conducted the ablation study to analyze the effect of the window shift operations. The results are reported in Table.~\ref{tbbb1}. The results indicate that the window shift operations can build cross-window connections and improve the performance of FTHNet.

\subsubsection{Depth and Patch Embedding Channel}
We explore the effect of different patch embedding channels $C$ and combinations ($N_1, N_2, N_3, N_4$) in the extraction backbone on model performance after choosing the BTB with W-MSA as the extraction block. 
The results are shown in Table.~\ref{tbdw}.

\begin{table}
	\footnotesize
	\centering	
	\caption{Ablation study of combination ($N_1, N_2, N_3, N_4$) and patch embedding channel($C$)}
	\scalebox{0.95}{
			\begin{tabular}{c c c c c c}
				\toprule
				 ($N_1,N_2,N_3,N_4$) & $C$ & SRCC & PLCC  & RMSE & Params(M) \\
				\midrule
				(2, 2, 2, 2) & 32 & 0.9328 & 0.9409 & 7.328 & \textbf{4.748}\\
                (2, 2, 2, 2) & 64 & 0.9355 & 0.9418 & 7.127& 11.70\\
                (2, 2, 6, 2) & 32 & 0.9331 & 0.9439 & 7.094& 5.558\\
                \textbf{(2, 2, 6, 2)} & \textbf{64} & \textbf{0.9358} & \textbf{0.9442} & 7.024& 14.88\\
                (2, 2, 6, 2) & 96 & 0.9332 & 0.9404 & 7.156& 30.34\\
                (2, 4, 6, 2) & 32 & 0.9303 & 0.9424 & 7.161& 5.662\\
                (2, 4, 6, 2) & 64 & 0.9338 & 0.9437 & \textbf{6.962}& 15.28\\
                (2, 4, 6, 2) & 96 & 0.9312 & 0.9430 & 6.971& 31.23\\
                (2, 2, 12, 2) & 64 & 0.9323 & 0.9415 & 7.112& 19.64\\
                (2, 2, 12, 2) & 96 & 0.9279 & 0.9395 & 7.039& 41.02\\
                (2, 2, 18, 2) & 64 & 0.9356 & 0.9416 & 7.266& 24.40\\
                (2, 2, 18, 2) & 96 & 0.9314 & 0.9374 & 7.2975& 51.71\\
				\bottomrule
	\end{tabular}}
	\label{tbdw}
\end{table}

It can be observed that our FTHNet-L (($2, 2, 6, 2$), $C$:64) can achieve the best performance. Compared with other patch embedding channels, 64 is the optimal choice. Considering the performance and model parameters, we finally chose (($2, 4, 6, 2$), $C$:32) as our FTHNet-S.

\subsubsection{Downsampling Structure}
We conduct this ablation study to explore the effect of downsampling structure in the Parameter Hypernetwork. Two downsampling structures are available in the Parameter Hypernetwork: stepwise and direct downsampling. 
As shown in Table.~\ref{tbbb1}, compared with the direct structure, the stepwise downsampling structure achieves almost the same performance with fewer model parameters and computational complexity. Thus, we choose the stepwise downsampling structure in our FTHNet.

\subsubsection{Loss Function}
We test different loss functions, including L1 loss, L2 loss, and Smooth L1 loss\cite{7410526}, while training the models for better performance. The L1 loss is the most common function in the IQA. In addition, L2 loss and Smooth L1 loss are also applied in the IQA. 

As listed in Table.\ref{tbbb1}, compared with other loss functions, Smooth L1 loss achieves the best performance in PLCC and RMSE. Meanwhile, Smooth L1 loss gets the second-best performance in SRCC. Thus, we choose the Smooth L1 loss as the loss function of our FTHNet.

\section{Discussion}

\subsection{Deployment Experiment}

We deploy our FTHNet as an Application Program Interface (API) on a GPU server and bind our FTHNet with a new automatic diagnosis system. This system can process fundus images and provide ophthalmologists with pathology information and the reliability of images. If the score of one image is high, the pathology information distinguished by the system shall be more trustworthy. The ophthalmologists utilizing this system can choose how to use the result based on the score of the FTHNet.

\begin{figure}[htbp]
\centering
\includegraphics[width=\columnwidth]{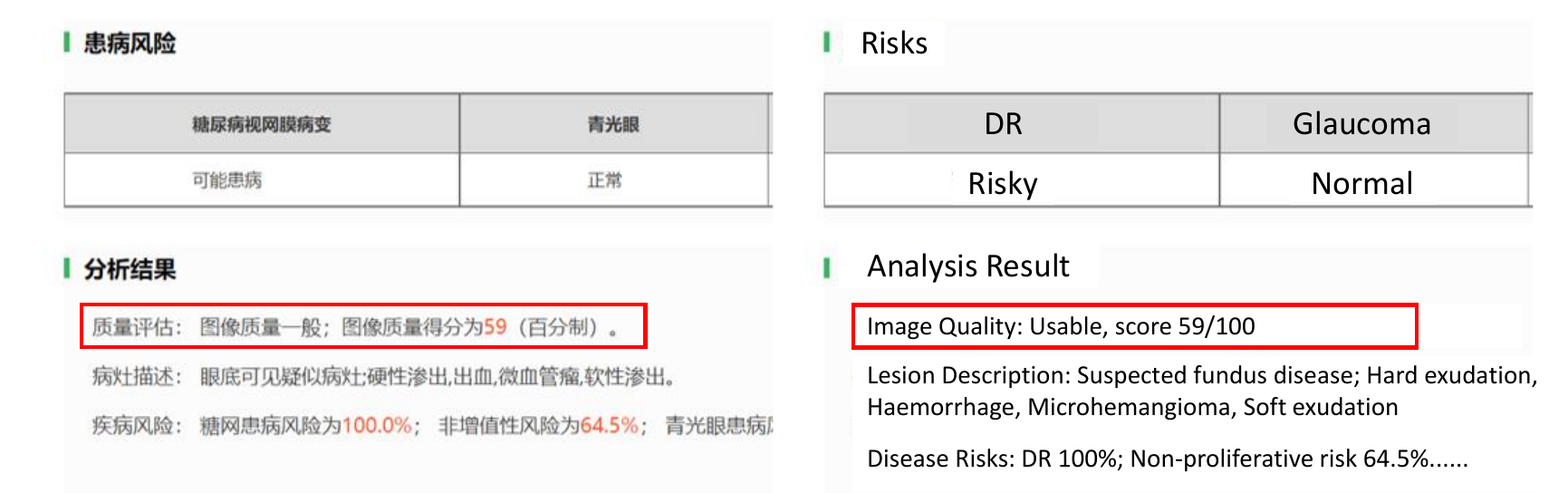}
\caption{Implementation of the FTHNet in the diagnosis system. The FTHNet works in the emphasized region.} 
\label{figempl}
\end{figure}

\subsection{Potential of Real-time assessment}
The inference speed is one of the most critical factors affecting the application of our FTHNet in clinical diagnosis. We deploy our FTHNet as an API on a GPU server to test the inference time of our FTHNet, and the results are in Table.\ref{tbrt}. The single test means the average inference time of a single fundus image.

\begin{table}
	\footnotesize
	\centering	
	\caption{Evaluation time consumption of different FTHNet models.}
	\scalebox{1}{
			\begin{tabular}{c c c c}
				\toprule
				 Methods & Params(M) & FLOPS(G) & Single Test(ms)   \\
				\midrule
			\textbf{FTHNet-S} & \textbf{5.662} & \textbf{1.981} & \textbf{44.45}\\
            FTHNet-L & 14.88 & 6.044 & 56.31  \\
				\bottomrule
	\end{tabular}}
	\label{tbrt}
\end{table}

We can observe from Table.\ref{tbrt} that the inference time of FTHNet-S is 44.45 ms, and even FTHNet-L is only 56.31 ms. With the short inference time, our FTHNet can provide real-time assessment in clinical diagnosis.

\subsection{Effect of hypernetwork Structure}
In this subsection, we discuss the effect of the hypernetwork structure on FIQA tasks.

To explore the importance of hypernetwork structure in the FIQA task, we conduct this experiment and show the results in Table.\ref{tbbb1}. Note that $w/o~hyper~network$ means the parameters of the Target Network are learned by backpropagation rather than provided by the hypernetwork.

It can be observed that the performance of FTHNet improves significantly with hypernetwork structure despite a slight increase in model parameters. It indicates the importance of hypernetwork structure in the FIQA tasks. Our FTHNet is constructed based on the hypernetwork structure. The Target Network has flexible parameters from the hypernetwork varying with the input image, which is essential for the quality assessment with fundus images of complex and diverse degradation.

\begin{figure}[!t]
\centering
\includegraphics[width=\columnwidth]{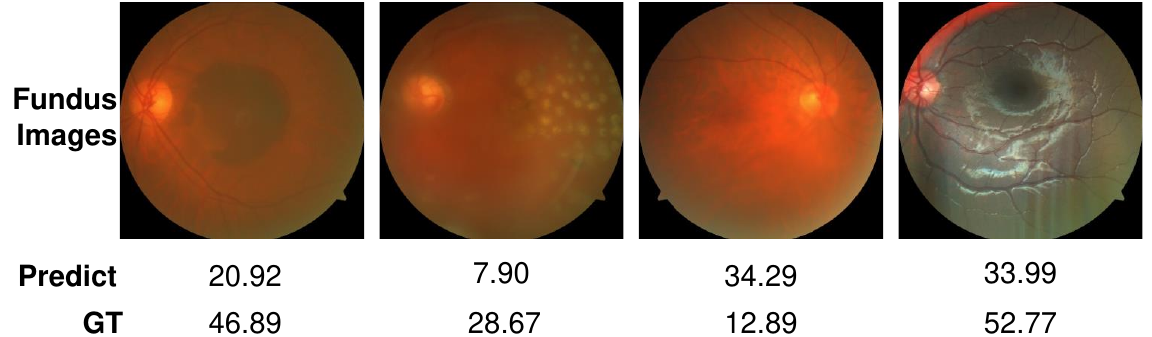}
\caption{Failed cases of FTHNet on our FQS dataset. Most of these failed cases are from the "Reject" category. The FTHNet and FQS dataset will be refined according to these failed cases.}
\label{figop}
\end{figure}

\subsection{Fail cases}
Although FTHNet achieves good performance, it may not work in some scenes. Fig.\ref{figop} shows some failed cases of FTHNet on our FQS dataset. As shown in Fig.\ref{figop}, there is a wide gap between the predicted MOSs and the ground truth. Further observation reveals that the three-level classification labels of these failed cases are all ``Reject". 
Furthermore, the degradation types are mainly a combination of severe blur and haze, with few samples in our FQS dataset. Though the prediction keeps them in the correct category, it shows that the number of samples limits the performance of our FTHNet in determining the scores for these images. We will continue to enrich and improve our FQS dataset according to these failed cases.

\section{Conclusions}

Traditional FIQA methods: most give classification prediction; it is difficult to compare results within each category; algorithms are old and not deployed in practice.

Our FTHNet: It is trained to give continuous quality scores, which are intuitive and convenient to compare; it is lightweight and easy to deploy in automatic diagnosis systems; it has a new transformer-based hypernetwork architecture and leading performance among IQA methods.

We will release codes of the FTHNet and full details of the FQS dataset for further research. Meanwhile, we are also willing to offer fundus image dataset-building tools for other fundus image research. We hope this work can serve as a baseline for FIQA and benefit the fundus image research by providing a new quantitative metric in the future.

\textbf{CRediT authorship contribution statement}

\textbf{Zheng Gong:} Conceptualization, Methodology, Software, Validation, Data Curation, Writing - Original Draft, Visualization. \textbf{Zhuo Deng:} Methodology, Software, Validation, Data Curation, Writing - Original Draft, Visualization. \textbf{Run Gan:} Conceptualization, Data Curation, Writing - Review and Editing. \textbf{Zhiyuan Niu:} Data Curation, Validation, Writing - Review and Editing. \textbf{Weihao Gao:} Validation, Writing - Review and Editing. \textbf{Lu Chen:} Data Curation, Project administration. \textbf{Canfeng Huang:} Data Curation. \textbf{Jia Liang:} Data Curation. \textbf{Fang Li:} Project administration. \textbf{Shaochong Zhang:} Funding acquisition. \textbf{Lan Ma:} Funding acquisition.



\textbf{Acknowledgments}

This work was supported by the Project from the Science and Technology Innovation Committee of Shenzhen-Platform and Carrier (International Science and Technology Information Center) and Shenzhen Bay Lab [KCXFZ20211020163813019]. This work involved human subjects in its research. Approval of all ethical and experimental procedures and protocols was granted by the Shenzhen Eye Hospital under the ETHICAL NUMBER 2022KYPJ062.

\appendix

\section{FundusMarking}

\begin{figure}[t!]
\centering
\includegraphics[width=\columnwidth]{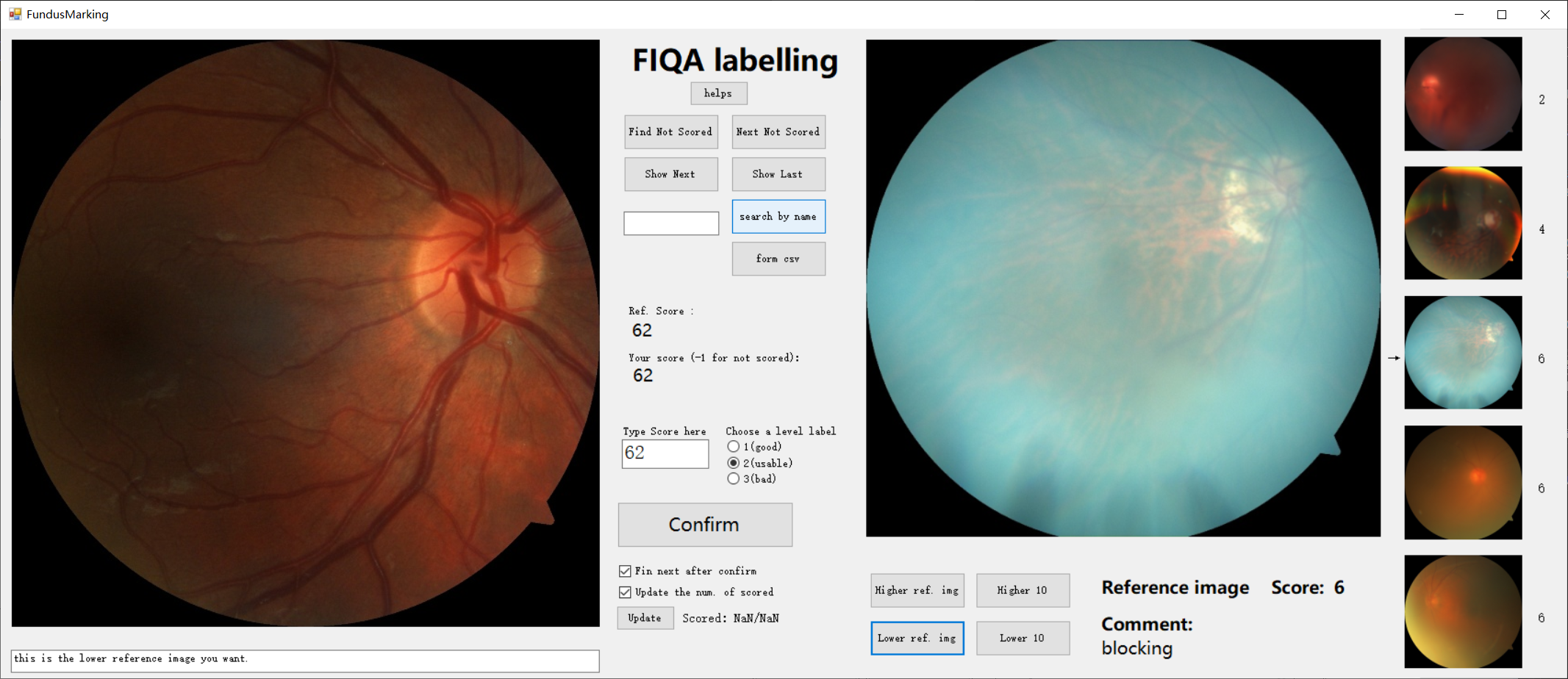}
\caption{The User Interface of our assistive labeling application, the FundusMarking. The FundusMarking is convenient and useful for inexperienced and experienced users, allowing ophthalmologists to provide normative source data easily. }
\label{figexe}
\end{figure}

Fig.~\ref{figexe} shows the User Interface of our assistive labeling application (FundusMarking).

\section{Downsampling structure in Parameter Hypernetwork}

\begin{figure}[htbp]
\centering
\includegraphics[width=0.8\columnwidth]{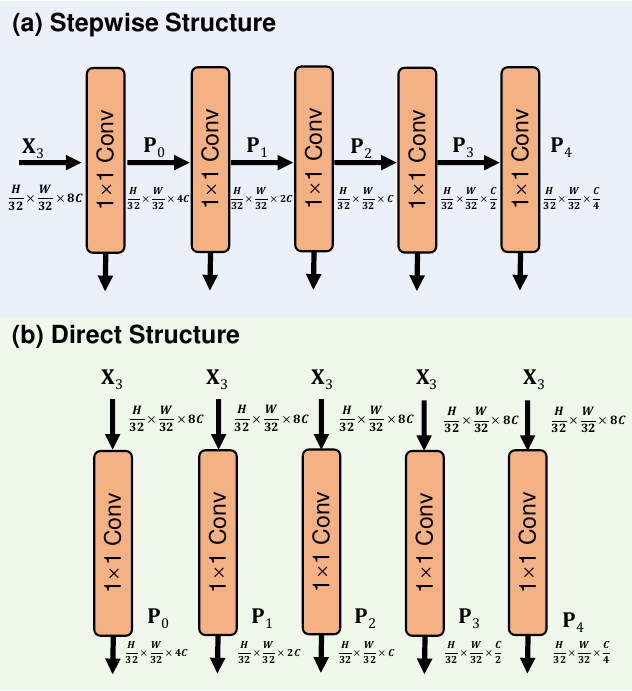}
\caption{The architecture of downsampling structure in Parameter Hypernetwork. (a) stepwise downsampling structure. (b) direct structure.}
\label{figshyper}
\end{figure}

As shown in Fig.~\ref{figshyper}(b), the direct downsampling of each stage gets the same input feature map $\mathbf{X}_3$ from the final BTB. In Fig.~\ref{figshyper}(a), the stepwise downsampling of each stage gets its input feature map from the previous downsampling layer, which utilizes the previous $1 \times 1$ convolution layer to reduce the parameters and computation. 

\begin{table}
	\footnotesize
	\centering	
	\caption{The theoretical parameters and computation of downsampling structure in Parameter Hypernetwork.}
	\scalebox{1}{
			\begin{tabular}{c c c c}
				\toprule
				 Structure & Layer & Params & FLOPS \\
				\midrule
				Direct & 1 & $\frac{1}{2}\times C^2\times K^2$ & $\frac{1}{2}\times C^2\times K^2 Map^2$\\
                Direct & 2 & $\frac{1}{4}\times C^2\times K^2$ & $\frac{1}{4}\times C^2\times K^2 Map^2$\\
                Direct & 3 & $\frac{1}{8}\times C^2\times K^2$ & $\frac{1}{8}\times C^2\times K^2 Map^2$\\
                Direct & 4 & $\frac{1}{16}\times C^2\times K^2$ & $\frac{1}{16}\times C^2\times K^2 Map^2$\\
                \midrule
                Stepwise & 1 & $\frac{1}{2}\times C^2\times K^2$ & $\frac{1}{2}\times C^2\times K^2 Map^2$ \\
                Stepwise & 2 & $\frac{1}{8}\times C^2\times K^2$ & $\frac{1}{8}\times C^2\times K^2 Map^2$ \\
                Stepwise & 3 & $\frac{1}{32}\times C^2\times K^2$ & $\frac{1}{32}\times C^2\times K^2 Map^2$ \\
                Stepwise & 4 & $\frac{1}{128}\times C^2\times K^2$ & $\frac{1}{128}\times C^2\times K^2 Map^2$ \\
				\bottomrule
	\end{tabular}}
	\label{tbpara}
\end{table}

Fig.~\ref{figshyper} shows the two different architectures of downsampling structure in Parameter Hypernetwork. Table.~\ref{tbpara} shows the theoretical parameters and computation of the downsampling structure in the Parameter Hypernetwork.



\bibliographystyle{elsarticle-num}
\bibliography{main_text}







\end{document}